\documentstyle [12pt]{article}
\evensidemargin\oddsidemargin
\begin{document}
\count0 = 1
\begin{titlepage}
\vspace{30mm}
 \title{\small{QUANTUM INFORMATION, IRREVERSIBILITY
 AND STATE COLLAPSE\\
 IN SOME MICROSCOPIC MODELS OF MEASUREMENT \\ }}
\author{S.N.Mayburov 
 \thanks{E-mail ~~ maybur@sgi.lpi.msk.su  ~~
 }\\
Lebedev Inst. of Physics\\
Leninsky Prospect 53\\
Moscow, Russia, 117924\\
\\}
\maketitle
\begin{abstract}
The  quantum measurement problem considered
for measuring system (MS) model which
 consist of measured state S (particle),
 detector D and information processing device O.
For spin chains and other O models the state evolution 
for  MS observables measurements studied.  
It's shown that specific O states structure forbids 
the measurement of 
MS  interference terms which discriminate pure and mixed
S states. It results in the reduction MS Hilbert space
to O representation in which MS evolution is irreversible,  
which in operational formalism  corresponds to S state collapse.
In radiation decoherence O model Glauber restrictions on 
QED field observables results in analogous
irreversible MS + field evolution. The results interpretation in Quantum
Information framework and
Rovelli's Relational Quantum Mechanics discussed.
\end{abstract}
\vspace{20mm}
\small{Talk given on 'Decoherence and entanglement' conference\\
Lake Garda , Italy, September 99\\}
\vspace {20mm}
\vspace{20mm}
\end{titlepage}
\section { Introduction}
 The fundamental problem of irreversibility
and in particular the state vector collapse in Quantum
Mechanics (QM) is still open despite the multitude of the proposed
models and theories ( for the review see $\cite  {Busch}$). 
 This paper analyses some microscopic dynamical models of quantum
measurements which attempt to describe
  the  evolution of the measuring system (MS) from
the first QM principles. In this models MS includes
the measured state (particle) S,
  detector D  amplifying S signal, environment E  and
 observer $O$ which process and store information. Under 
observer we mean information gaining and utilizing system 
(IGUS) of arbitrary structure  $\cite {Gui}$.
 It can be both human brain and some automatic device 
processing the information , but in both cases it's
the system with many internal degrees of freedom (DF) which permit to
memorize the large amount of information.
In general the information processing or perception in case of human brain
 is the physical objects evolution which on microscopic level
supposedly obeys to  QM laws   $\cite {Alb}$.
For example, the information bit transfer in standard computer
 corresponds to the electrons motion inside the
semiconductor chips. Such process  can induce the back-reaction  
on the information gained in the measurement
 and must be accounted  in the  general measurement theory.
  In this paper we'll use the brain-computer analogy without
discussing its reliability and philosophical 
implications $\cite {Pen}$. We'll ignore here quantum computer options
 regarding only standard solid-state dissipative computers.

The possible role of observer  
in the wave-function collapse was discussed for long time \cite {Wig}, but 
now it attracts the significant attention again  due to the
the progress of quantum information studies $\cite {Pen}$. The
 formal description of standard QM and its generalization
for different observers in terms of Information Theory 
was developed by Rovelli 
 $\cite  {Rov}$. We can't describe here this
formalism called Relational QM at length citing only the features 
essential for quantum measurements. In particular its definition
of quantum information via correlations between  S and $O$ states
will be used throughout the paper.

Standard Copenhagen QM interpretation divide our physical world
into microscopic objects which obeys to QM laws and macroscopic objects
, also observers which are strictly classical. This artificial partition
was much criticized, first of all because it's not clear where
to put this quantum/classical border. Moreover there are strong
experimental evidences that at the dynamical level no such border
exists and QM  successfully describes large, complicated systems
including biological one.    
Following this conclusions Relational QM  concedes
(Hypothesis 1 of Rovelli paper) that QM description is applicable
both for microscopic  states  and macroscopic
objects including observer $O$ which  Dirack state vector
$|O\rangle$  can be defined relative to  some other observer $O'$, 
which is also another quantum object.
 The evolution of any complex system C
 described by Schrodinger equation of some
(may be very complicated) form and for any  C
including MS the superposition principle hold true 
at any time. Let's consider in this ansatz  $O'$ description of
 the measurement by $O$ which starts at $t=t_0$
of binary observable $\hat{Q}$ on  $|s\rangle=a_1|s_1\rangle+a_2|s_2\rangle$
, where $|s_{1,2}\rangle$ are $Q$ eigenstates.
 It follows from the linearity of Schrodinger equation
that for $t>t_1$ when $O$ finished to measure S state
 the state of MS system  relative to $O'$ observer is
\begin {equation}
   \Psi_{MS}=a_1|s_1\rangle|D_1\rangle|O_1\rangle+
a_2|s_2\rangle|D_2\rangle|O_2\rangle
                                   \label {DD}
\end {equation}
Here $|O_{1}\rangle$ is $O$ state vector
 after finishing the measurement of particular $S$
state  $|s\rangle=|s_1\rangle$ (and  correspondingly for $s_2, O_2$).
 After S-D-$O$  interaction (measurement) finished and  $O$ internal
state changed, it means  $O$ memorizes the quantum
information  about $Q$ value for S state. 
 For example $|O_{1,2}\rangle$ can correspond to some
excitations of $O$ internal collective DF
 like phonons, etc., which conserves this information.
 For the simplicity in the following  we'll  omit detector D in 
measurement system MS
assuming that S directly interacts with $O$. It's reasonable for simple models,
because if to neglect decoherence the only  D effect is
the amplification of S signal to make it conceivable for O.
In our formalism we'll use C general relative pure states
$|C^i_j)$  which in principle can differ from Dirack vectors.
 Here the lower index marks C parameters 
and upper one - observers ( for $O,O'$ they are $0,1$),
 for example  $\Psi_{MS}=|MS^1)$.

  Relational QM 
formalism by itself can't resolve the state collapse enigma
but give us some additional insight from the comparison
of $O, O'$ observers reports on    S measurement result. 
In this approach at time $t>t_1$  for observer $O'$
MS is  in the pure state  $\Psi_{MS}$ of (1).
Yet we know from the experience that 
 at the same time $t$  $O$ percepts the final
MS state as the mixed state $\rho_m$, which 
means the state collapse. Considering this paradox
we must notice that MS measurement by $O$ 
formally  includes the measurement
of $O$  own internal DF or selfmeasurement $\cite {Bre}$.
In relation with it Ashtekar proposed
 phenomenologically  that for observer $O$ some of its own
internal DF $Q_U$ can be principally unobservable
($\cite {Rov}$ and ref. therein).
 If so then for observer $O$
 the result of selfmeasurement
  described  taking the trace over $Q_U$ and due to it
  $O$ subjective state $|O^0)$  always percepted as mixed one.
If this state entangled with S  state after S measurement
then the complete MS
state is also looks mixed for $O$, but not for $O'$.
To investigate this promising  idea it would be
 interesting  to look for this
unobservable DFs in some  measurement models.
 
The selfmeasurement problem  often regarded as the implication
of more general algebraic problem of selfreference $\cite {Busch}$.
Following this approach Breuer has shown
 that the phenomenological selfmeasurement 
restrictions for classical and quantum measurements are
analogous $\cite {Bre}$. Yet at least in quantum case they are
introduced ad hoc and don't obtained from Schrodinger linear MS evolution.
In fact Breuer formalism needs additional QM collapse  postulate
   in a weaker
'subjective' form. In our approach  we'll follow different route
 assuming that MS evolution
in selfmeasurement is also linear and obeys to
Schrodinger equation of some form.  Then we'll try to find
the restrictions  and unobservable DFs  in some microscopic
IGUS  measurement models. In chap. 2 it will be shown that
Heisenberg commutation relations  for $O$ operators and $O$ particular
 atomic structure  restricts the information
acquisition and results in collapse-like evolution.
Decoherence effects  can be important also in IGUS 
selfmeasurment models \cite {Zur}. To check it in chap.3
we'll consider the particular mechanism of MS radiation decoherence
which applied   for the information 
memorization  in the realistic IGUS models.
In chap. 4 we'll discuss the physical and phylosophical implications
of described models results.
   
To relate our models with Measurement Theory we'll introduce
strict measurements defined as follows : for  observable $Q$
defined on $h_s$ - S subspace of MS Hilbert space,
 exists $Q_o$  on $h_o$ - $O$ subspace,
for which $\bar{Q}=\bar{Q}_o$. The same definition
 can be used for detector D, operator $Q_D$
and its subspace $h_D$. The
 difference operator $\Delta_Q=Q-Q_o$ defines measurement uncertainty
and estimate of information available for $O$ on $Q$ value.
If $\bar{\Delta}_Q=0$ this is exact measurement transferring maximal
information on $Q$ value for S state. 
For the case when only
 detector D considered in the model the strict measurement
  in fact describes any realistic
experiment  $\cite {Busch}$. This is close analog of the
 measurement of first kind,
but if observer also
included in the model some new features  appears, as will be shown below.

The formal selfmeasurement definitions have some distinctions
(compare \cite {Alb} and \cite {Bre}), but for our models they
are inessential. As the example of selfmeasurement let's 
consider some automata $O$ which consist of ensemble of 
independent binary pointers
 $|G_{i}^j\rangle, i=1,2, J=1,N $. Each pointer can
 interact with
external state S or other pointer and 'measure' their states analogously to
 eq. (1). If some $O$ program control pointers $G_j, G_k$ interaction sequence
then $O$   performs selfmeasurement $\cite {Alb}$.   

In our  models we'll suppose that MS  always can
 be described completely (including E if necessary)
 by some state vector $|MS\rangle$ relative to 
$O'$. MS can be closed system , like atom in the box or open pure
system  surrounded by electromagnetic vacuum or  E of other kind.   
Throughout this paper the operational definition of
collapse used : if any (self)measurement performed by $O$ can't indicate
the difference between mixed and pure MS state then this state
percepted by $O$ as mixed one.
 The results of measurements in this approach
estimated and compared by $\bar{Q}_O$ value of any true  observable $Q_O$.
This definition isn't complete by itself and can't account the 
'problem of event' directly $\cite {Busch}$, but for models study
it's quite useful.

\section {Quantum Information and IGUS Model}

Till now there are only few attempts to construct the observer
microscopic models which can make discussion more fact-like
$\cite {Don}$. Here we consider the simple microscopic IGUS model
of  information processing and memorization
which reveals some of its important features.
IGUS  selfmeasurement restrictions must be
  defined by IGUS  physical structure, in particular 
its atomic structure. Due to it even macroscopic observer can
store only finite amount of information and this
IGUS finiteness will be shown to have important consequences for
quantum information acquisition. Moreover it's reasonable to
demand that the (self)measurement process
 shouldn't destroy IGUS 
structure or obstacle to its proper functioning
and conserve previously stored information.
It's reasonable to express this conditions via structure
conserving operator
$\hat{R}=\prod \hat{P}_j$ , where $\hat{P}_j$ are some projection
operators describing  IGUS structure. All the physically
consistent IGUS states must be $P_j$ eigenstates and
this conditions should be fulfilled also during and after the measurement. 
For example
if IGUS has the atomic crystal lattice, then
 projector $\hat{P}_r$ permits only limited
range of  distances between neighbor atoms $|r_{i,i+1}|<a_0$
in IGUS state vector.

We consider first the toy-model  based on Coleman-Hepp (CH) model  which
 used  often for QM paradoxes discussion $\cite {Hep}$.
CH model considers  fermion $S^0$ spin z-projection measurement via
interaction with $N$ spin-half atoms $A_i$ linear chain -
1-dimensional crystal detector  D. $A_i$
atoms are regularly localized at the distance $r_0$ by the effective potential
 $U_i(x_i)$. $S^0$ initial 
state $\psi^0_0=\varphi(x,t_0)(a_1|u_0\rangle+a_2|d_0\rangle)$ 
where $u,d$ are up,down spin states and $\varphi(x,t_0)$ is localized $S^0$
 wave packet spreading along  D spin chain. 
For the comparison  the measurement of corresponding
 mixed state $\rho^0_m$ with   weights $|a_{1,2}|^2$ will be regarded.
 $S^0 - D$ interaction Hamiltonian  is:
\begin {equation}
   H_I=(1-\sigma^0_z)\sum^N_{i=1} V(x-x_i) \sigma^i_x  \label {A0}
\end {equation}
where $V$ is $S^0-A_i$ interaction potential.
 For suitable  model parameters and for
 D initial polarized state  $\psi^D_+=\prod|u_i\rangle$ one obtains that
 if $S^0$ initial spin state is $|u_0\rangle$ this D state conserved
 after $S^0$ passed over the  chain, 
but for initial state $|d_0\rangle$ 
D state transformed into $\psi^D_- =\prod|d_i\rangle$. 
Thus for  finite $N$ at $t>t_1$ for $S^0-D$ final state
\begin {equation}
 \psi_f(t)=\psi_1(t)+\psi_2(t)=
\varphi(x,t)(a_1|u_0\rangle\psi^D_++a_2(-i)^N |d_0\rangle\psi^D_- )
                                  \label {A2}
\end {equation}
we get macroscopically different values of
D pointer which described by the polarization operator  :
 $\mu_z=\frac{1}{N}\sum \sigma^i_z$ acting in $h_D$ subspace.
It gives estimate $\bar{\mu}_z=\bar{\sigma}^0_z$ and 
$\bar{\Delta}_Q=0$, so this is strict exact
measurement.  
Despite, it doesn't mean $S^0$ state collapse because $S^0 - D$
interference terms (IT) operator:
\begin {equation}
B=\sigma^0_x B_I=\sigma^0_x\prod_{i=1}^N\sigma^i_y   \label {A2A}
\end {equation}
describing spin-flips of all $A_i$ and $S^0$ spins.
In principle $B$ also can be measured by
observer $O$ and discriminate $S^0 - D$ mixed and pure states.
Its expectation value $\bar{B}=.5(a_1^*a_2+a_1a_2^*)$
 for $S^0 - D$  final state $\psi_f$ differs from
$\bar{B}=0$ for $S^0$ mixed  state $\rho^0_m$ $\cite {Bel}$.
For the convenience we exclude from consideration
 $a_1, a_2$ values such that $\bar{B}=0$,
which doesn't influence on our final results.
 Note that $S^0$ IT can be measured separately, but
 only before $S^0-D$ interaction starts, after it only their joint IT operator
have sense.
 $\mu_z,B$ don't commute and can't be measured
simultaneously:
\begin {equation}
  [\mu_z,B]=\frac{i\sigma^0_x}{N}
\sum_{i=1}^N \sigma^i_x\prod^N_{i\neq j}\sigma^j_y
            \label {A3}
\end {equation}
It's easy to propose how to measure collective (additive) operator $\mu_z$,
 but also 
$B$ values can be destructively measured  decomposing D into atoms
and sending $A_i$ one by one and also $S^0$ into Stern-Gerlach magnet.
 Then measuring
$A_i$ amount in each channel and their correlations by some other detector
D$'$ one obtains information on $B$ value from it.
Note that $B$ measurement isn't strict for D subspace,
but is strict for $D'$, the fact which will be used below.
Standard QM don't regard any special features of
 destructive measurements assuming that any hermitian operator
is observable and can be measured by one way  or another.

 Of course  CH model only crudely imitates 
 the evolution of real detectors which are the collective
solid states  characterized by  the strong atoms selfinteraction.
In CH model  it's accounted only by effective potential $U_i$ and don't
depend on $A_i$ state.
If one wish to consider closely the effects of atoms selfinteraction
on the interference terms then
for CH model it can be simulated by neighbor spins
interaction adding to $H_I$ Heisenberg Hamiltonian 
describing ferromagnetism :
 $H_f=J(r_{i,i+1})\sum\vec{\sigma}^i\vec{\sigma}^{i+1}$
, where $J$ is spins interaction potential called also exchange
integral.
Note that our D initial state $\psi^D_+$ is  $H_f$ eigenstate
and the excitations of this states by $S^0$ for large $N$ are
described by spin waves - magnons $\cite {Zim}$. This changes our picture
only in this sense that now the operators describing
D state - $\mu_z, B$ must be changed to some time dependent
operators , but commutation
relations still are the same and all our conclusions
will be true for them. The more realistic imitation of real
detectors seems the
 electron excitations  and  the  lattice excitations - phonons
  which dynamics reminds
the magnons one very closely $\cite {Zim}$. Note that due to
selfinteraction the studied D state will be largely perturbed
if one tries to decompose D into atoms for IT operator $B$ measurement
$\cite {May}$.

From analogous  considerations some authors assumed 
that for collective selfinteracting  systems D IT operators are unobservable
and so can explain the collapse  $\cite {Hep}$.
We don't regard this hypothesis as well founded and at least
it needs consistent proof of such restrictions existence, analogously
to described IGUS restrictions. 

 Extending this ideas to construct the observer or IGUS
models one should take into account 
that observer  performs the selfmeasurement of its own state
and so obeys to additional $R$ restrictions..
 For example let's consider that  
the input  signal  induces inside solid state IGUS the
electron loop current state $|\psi_J\rangle$, which stipulate
 signal perception.
The electron currents in brain or processor in general are very complicated,
 but  for the simple model  of binary state measurement we  assume
that one of this states -  $|d_0\rangle$  excite in the conducting loop
 the current and other one - $|u_0\rangle$ not. Then
neglecting fermion statistics the
current operator have additive form $\vec{J}^p=\sum e\vec{v}_i$, where
$\vec{v}_i$ is the electron velocity operator, which for this binary
signal has only eigenvalues $0$ and $\vec{v}$. So in first approximation
 the current operator structure  is analogous to CH model $\mu_z$
polarization
and we can study distinction between the pure and mixed states for it 
to understand some effects for current states.

For this purpose let's consider ensemble of $N_c$ CH spin chains  as our IGUS
 $O$ performing $S^0$ spin measurement and start with $N_c=1$. 
In CH model the information about $S^0$ spin acquired and memorized by  $O$
 in the form  of observable $\mu_z$ eigenstates  superposition of (\ref {A2}). 
Normally  IGUS consists of acquisition (perception) channel AC which interact
with S , transfer and amplify its signal
and memory cells MC which interact with  AC and memorize
information about S. In CH model  we formally can  regard
first $N-1$ atoms as our AC and $A_N$ as single MC for which
$\bar{\sigma}^N_z=\bar{\sigma}^0_z$.
But for simplicity we'll consider  the spin chain as AC and MC 
simultaneously, so MC subspace for $N_c=1$ is $h_O$. 
 Memorization by MC can be only
strict measurement, alike $\mu_z$ memorization in $h_O$. It's impossible
 for $O$ to 
memorize $B$ value because it isn't operator on $h_O$ only. 

 The real difference of $S^0 - O$ measurement by single chain from
 $S^0 - D$ measurement  described above  is that
for given $O$ structure it is principally impossible to perform
 $B$ selfmeasurement by $O$.
 To demonstrate it
let's compare   $S^0 - D$ interaction observed by $O'$
with   $S^0 - O$ interaction  observed by same $O'$.
In first case after the interaction finished the observer $O$ had the
 choice to percept and memorize 
 $\mu_z$ or to decompose D into
atoms and measure $B$ and to verify that the state is pure or mixed.
  In the second  case when  $O$ finished to interact
with $S^0$ pure state and  memorized $\mu_z$  for
 $O'$ the system
$S^0  - O$ final state is also $|(S^0-O)^1)=\psi_f$ of (\ref {A2}).
 We settled preliminarily that $O$ state should obey to $R$ restriction and
in particular $P_r$ which means that $O$ spin chain can't be
decomposed into atoms. 
 Yet if $B$ measurement as we assumed  can be only destructive for 
$O$ and destroy its chain structure then after it $O$ simply
stops to function and can't memorize $\mu_z$ or $B$ information.
 So we conclude that for properly  functioning observer $O$ operator $B$
, or more exactly $B_I$ 
is principally unobservable. Consequently it changes $O$ state space
 so that there is no difference
between pure and mixed $S^0 - O$ states. 
 Roughly speaking there is no Stern-Gerlach magnet in the brain and if to
 decompose the brain
into atoms and send them to external analyzer the brain will broke 
unrestorably and stop to function.
 Of course this operation - $B$ measurement
can be performed by external observer $O'$ and verify that MS state is
$\psi_f$ but due to $O$ destruction in this process $O$
never can obtain this information from $O'$. 
In our opinion obtained paradox shows that in $O$ basis the operator $B$ 
is unobservable and $|(S^0-O)^0)$ final states in
selfmeasurment formalism are effectively the same
for pure and mixed initial $S^0$ state.

Our selfmeasurement formalism corresponds with Heisenberg
or algebraic QM, where states manifold defined by observables set,
so if some hermitian operator $B$ is unobservable it reduces
 initial Hilbert space
to some new $O$  states manifold.
 To describe $O$ states manifold we'll use  Hilbert space $H'$ of
observer $O'$ which include all the possible physical
 states $|\psi^1\rangle$ of surrounding world, except 
$O'$ internal states. So assuming external-internal states
factorization the complete Univers Hilbert space is
$H_T=H'_e*H'_i$ tensor product, where $H'_i$ is $O'$ internal states subspace.
 $H'_e$ space is spanned
on Hermitian operators $Q'$ which are $O'$ observables set. For observer  $O$
all its Hermitian operators $Q$  ( except operators describing
$O'$ internal DF) can be obtained mapping $Q'$ to $O$ rest frame 
$Q=UQ'U^+$. From the external-internal states factorization
it follows 
 $H_T=H_e*H_i$, where $H_e$ is  Hilbert space
of outside world and $H_i$ is Hilbert subspace of
$O$ internal states. In  particular $O'$ observable $B'$ transforms into $B$
which as we supposed is unobservable for $O$.
Due to it $M_I$ - the space of $O$  internal states $|O^0)$ is nonequivalent
to $H_i$, but can be decomposed as the tensor product of 
Hilbert subspaces $\cite {Busch}$.
 Each of this subspaces consist of $O$ eigenstates
of given measurement spectral decomposition ( for example $\mu_z$ eigenstates
for CH model)
 of the same particular eigenvalue : $M_I=h_1*h_2...h_n$.
Unitary states transformations from $h_l$ to $h_j$ performed by unobservable
 $B$ only and so such unitary transitions are unphysical.
 Eventually the total $O$ states $|\psi^0)$ manifold is
 $M_T=M_I*H_e$ relative to 
$O$ observer and it describe the observer representation
or observer basis $\cite {Zur}$. In this basis
$|MS^0)$ states coincide with correspondent mixed states
$\rho^{ij}_m=\delta_{ij}|\psi_i\rangle \langle\psi_j|$
for $\psi_i$ of (\ref {A2}). It means that at any time for any
true observable $Q$ we have:
$$
     \bar{Q}=Tr  \rho_p Q =tr  \rho_m Q
$$
where $\rho_p$ is corresponding pure state and this result corresponds to  the
 state collapse in operational approach.

So due to our R restrictions and $B$ nonobservabilty for the interacting
 observer $O$ its own states Hilbert space is  
 reduced in comparison with $O$ states space relative to
  observer $O'$ noninteracting both with $S$ and $O$. $O'$ can describe
 their evolution   by Schrodinger equation but $O$ can't do it  starting
 from the moment
$t_0$ when $S$ starts  to interact with $O$. Consequently $|MS^0)$ evolution
for $O$ observer coincide with $\rho_m(t)$ evolution which
describe the stochastic quantum jumps from initial to one of final
MS states.  
 Due to the  appearance of $S^0, O$ states entanglement after
 $S^0 - O$ interaction in CH model
 $O$ representation of MS final states is also equal to tensor
product of subspaces characterized by $\mu_z$ eigenvalue which have each
vector of this subspace i.e. it is mixed state corresponding to $\mu_z$
 measurement and collapse.

 For the collective
solid states with strong selfinteraction 
the evolution also changes as was discussed above for D spin-spin
interactions which
probably makes R observation restrictions even more stiff.
In the current loop model to measure corresponding $B$ value
means to measure $N$ electrons interference inside crystal
with which this electrons also interact strongly.
 It's the same kind of restrictions like in CH model, but
it's even more obvious that
nondestructive $O$ selfmeasurement is impossible.  
In the next chapter it's shown that atoms selfinteraction effects
can be described by QED radiation decoherence model which results in
quite different restrictions.

Of course this consideration isn't consistent proof of
$B$ nonobservability for observer $O$. Rather it demonstrates
 qualitatively that for realistic collective $O$ systems IT selfmeasurement
  necessary for the collapse verification
obstacles effective information processing and memorization by $O$.
In particular we can't strictly prove that $B$ measurement can be only
destructive, but for any realistic measurement Hamiltonians it's
impossible to perform it otherwise for large $N$.

Now we'll argue that the analogous effects can be derived in some cases
even without assuming $B$ measurement to be destructive.
Comparing our model results with Ashtekar hypothesis it seems that 
$O$  unobservable operators aren't particular DFs, but IT which
can be constructed of any DF combinations and they defined in fact
by $O$ operators which are measured in the particular experiment
 and their commutation relations. Let's consider their connection
with $O$ structure and maximal available information.
For this purpose we'll use  CH observer model and 
suppose that $O$ consist of $N_c=m$ chains denoted $D_c, D_c',...D_c^m$
and their subspaces are $h_c, h'_c,...,h^m_c$. $O$ structure defines
$S^0-D_c$ interaction Hamiltonian $H_I$ of (\ref{A0}).
Due to its asymmetry relative to $h_S,h_c$ 'axes'
the information on $\sigma^0_z$ transferred
exactly to $h_c$, but $\sigma_x^0$ information corresponding $S^0$ IT
don't transferred at all. As the result after $S^0-D_c$ interaction finished
at $t_1$ and they parted no strict measurement by $O$  
 can discriminate pure or mixed $S^0$ states.
But it can be done by the operator $B$ measurement by the next chain.
To perform it at  $t>t_1$
  $S^0$,$D_c$ must interact via some Hamiltonian $H'_I$ 
with  $D'_c$ which is also $O$ part. $D_c$ state at $t>t_1$
is $\psi_f=b_1|B_+\rangle+b_2|B_-\rangle$ of ($\ref {A2}$)
rewrited here as sum of $B$ eigenstates and so the complete $O$ state
before $H'_I$ interaction at $t=t'_1$ turned on is :
$$
\Psi_{MS}^1=\varphi'_0=\psi_f|D'_{c0}\rangle...|D^m_{c0}\rangle
$$
Then after this $S^0,D_c,D'_c$ interaction finished at $t=t_2$ $O$ state 
becomes :
$$
 \Psi_{MS}^1= \varphi'_f=\varphi_1'+\varphi'_2=
(b_1|B_+\rangle|D'_{c+}\rangle+b_2|B_-\rangle|D'_{c-}\rangle)|D^2_{c0}...
|D^m_{c0}\rangle
$$ 
After it $D'_c$ contains information on ${B}$ expressed
 by strict operator $B'$ in $h'_c$ for which $\bar{B}'=\bar{B}$.
Yet in the same time $\sigma^0_z$ information disappears in $h_c$ where
 $\bar{\mu}_z=0$ at $t>t_2$ in distinction from true
  $\bar{\sigma}^0_z$ value which was kept by $D_c$ till
$t<t'_1$ . So the full information about
S state never acquired by $O$, due to $B,\mu_z$ incompatibility. 

Yet $B$ measurement by $O$ doesn't mean that $O$ can discriminate
its own (or MS) pure and mixed state. To perform this discrimination
 the next chain $D^2_c$ 
should measure $S^0,D_c,D'_c$ state $\varphi'_f$ joint IT operator
 $B_2$ which form is analogous to $B$ of (\ref {A2A}). It can 
be presented as the operator sum of $N+1$ members: 
$$
     B_2= \prod_{j'=1}^{N} \sigma_y^{j'}\sum_{n=1}^{N+1} B^p_n
$$
where $j'$ means $D'_c$ elements array, and sum members are :
$$
  B^p_0=\sigma^0_z  ; \quad B^p_1=\sum_{i=1}^N \sigma^i_x ; \quad 
  B^p_2=\sigma^0_z\sum_{i=1}^{N-1}\sum_{l>i}^N \sigma_x^i \sigma_x^l  ;\quad
  B^p_3=\sum_{i=1}^{N-2}\sum_{l>i}^{N-1}\sum^N_{k>l}
 \sigma^i_x\sigma^l_x\sigma^k_x
  ; \quad ...
$$
and so on to include all uneven  number spin flip combinations.
This $B_2$ measurement  gives IT estimate between $B$ eigenvectors
 $\varphi'_{1,2}$, 
but as the result the new entangled state $\varphi^2_f$ appears
 including in addition $D^2_c$, which IT also must be measured to
define MS state.
To measure full $O$ state ITs this measurement sequence should be
extended till it includes $N_c=m$ chain, but its IT can be measured only by
external $O'$. So due to $O$ finiteness there is always at least
one MS IT operator $B_m$ unobservable for $O$.
The real meaning of this result needs further study, and here we
propose only its tempting interpretation.
 In our opinion it evidences that even if
any $B_j$ measurement can be nondestructive nevertheless
, due to $O$ finite rigid structure its selfmeasurement
don't permit to discriminate pure and mixed MS states $|\Psi^0_{MS})$.  
For example for $N_c=2$ operator $B_m=B_2$ can be measured only by $O'$
for which MS state is $|\varphi'_f\rangle$, but   $B_2$ is
unobservable for $O$  and can't discriminate MS pure and mixed states.      
It's not clear to which extent this result can be general
 $O$  selfmeasurement property,
but we notice that operator $B_m$ acts on all $O$ DFs and to
measure and memorize $B_m$ we need at least one more DF which can't belong to
$O$.

\section {Radiation Decoherence (RD) and IGUS Memorization}

Here we'll discuss  the decoherence model of IGUS which imitate
information processing in elementary computer $O$  
memorizing quantum signal. 
 We'll assume that its structure
is the finite regular monoatomic crystal lattice L. 
In this case the   lattice excitations and
dislocations can store the quantum information.
 L initial environment $E_f$ is taken to be the electromagnetic vacuum in
its ground state $|V_0\rangle$.
As the example we'll consider the inelastic collision
of S - neutral  particle $n$ wave packet with L in  ground state $|L\rangle$.
 If one of $n$ trajectories $x_2$
crosses L aperture and other $x_1$ lays outside it, such MS
 can perform the measurement of $n$ position $x$.
Suppose that this impact starting 
at $t_0=0$  produces excited state $|L^*\rangle$ which
produce  dislocation with the new ground
state $|L'\rangle$ and accompanied by  multiple  
phonon excitations. This excitations  can be dissipated completely
via cascade decay of phonons into photons $p\rightarrow p'+\gamma$,
 which detailed dynamics described in
$\cite {Zim}$. So at  large time the lattice transferred to new ground state 
L$'$ and all energy excess dissipated to electromagnetic field.
 Then at large time $t\rightarrow\infty$ the final state of
our system and environment becomes :
\begin {equation}
   \phi_f(t)=\phi_1+\phi_2=a_1 |x^n_1\rangle|L\rangle|V_0\rangle +
a_2\sum c_j|x^n_2\rangle|L'\rangle|j \gamma\rangle \label {C1}
\end {equation}
where $|j \gamma\rangle$ is the localized  state of j photons (packets)
 orthogonal to $|V_0\rangle$, and $c_j$ are their production
amplitudes. So $O$ stable states L,L$'$ memorize
information about $x^n$. 
 The analogous final state is produced
if $n$ impacts and excites the single molecule $L\rightarrow L'$
memorizing information, which model will be described in forcoming paper.

Beside the possible nonobservability of $L, L'$  interference 
discussed in the previous chapter the analogous effect can be found
 for the decohering electromagnetic field $E_f$. QED field measurements
are described by Glauber photocounting theory confirmed now
experimentally. In its framework
  all the field observables which can be measured by material
detectors $D_{\gamma}$  are
the algebraic functions $F$ of photon numbers operators 
$\hat{n}(\lambda, \vec{k})$ only $\cite {Gla}$. But any hermitian
operators witn nonzero matrix elements between the states with different 
photon numbers like $V_0$ and $j \gamma$ can't be such functions
and so their interference is unobservable directly 
(no interference with vacuum !). So for any true $E_f$ observable 
$Q_E=F(\hat{n})$ we have : 
\begin {equation}
 \langle V_0 |Q_E|j\gamma\rangle=0 \label {C2}
\end {equation}
and its effects coincide with mixed state ones. As the result
for any true observable $Q$ on $n,L,E_f$ space it follows :
\begin {equation}
                     \bar{Q}=Tr \rho'_p Q=Tr \rho'_m Q \label {C22}
\end {equation}
where $\rho_{p,m}'$ are the corresponding pure and mixed density
matrices,   $\rho_m=|\phi_{i}\rangle\langle\phi_i|$ of (\ref{C1}).
 Consequently not only  observer $O$ identified
with L , but also any other $O'$ can't discriminate the pure
and mixed initial $n$ states in this case.
Even if one admits that L, L$'$ (and $x^n_{1,2}$)
IT  operator $B$ can be measured 
and any information from $D_{\gamma}$ accounted by $O$,
 it's impossible to measure
the interference terms for the complete state $\phi_f$ of ($\ref {C1}$).

 Note that the initial vacuum state can include
arbitrary number of photons noncorrelated with $L$ state and
it doesn't change the final result. Despite that this model
is  oversimplified it can have some relation
to the real computer or brain, where the input quantum  signal induces the
motion of electron currents. This electrons scattered by the crystal
lattice excite it and produce via phonon decays soft (thermal)
radiation. In the current loop model described
 in   chap.2 , the final state can be analogous to (\ref {C1}) 
, if electron pulses coupled to some stable states
 L,L$'$ like ferrite memorizing rings. 
 So this information processing and memorization induce the energy
dissipation or decoherence, analogously to results of 
 classical information and selforganization theory $\cite {Gui,Lan}$.
In general such $L$ evolution seems quite typical for measurement.
First to shift our 'pointer' some energy is needed stored in
detector metastable state or particle $n$ energy.
Then this energy must be dissipated for information memorization in some 
$O$ ground states.

The well-known argument is
that by means of suitable mirrors array the produced radiation
can be reflected,
reabsorbed and  the initial n-L state restored  $\cite {Buc}$.
This is in fact only approximately true, because absolutely
reflecting mirrors are prohibited by the laws of physics and
so at slow rate the radiation always penetrate them.
It means that even if our IGUS contained inside the mirror
box  it will function, but more slowly and ineffectively.
 In  case of real mirrors
this mirror box must be accounted in the quantization as MS part
and we must analyze $E_f$ field penetrating through its walls. 
 Then  the multiphoton states of this newly quantized free
field $E_f$ outside of box must be regarded. It's reasonable to
 suppose that asymptotic
MS states at $t\rightarrow\infty$ will coincide also with
 (\ref {C1}), so our IGUS will function, but in
different regime. If we regard IGUS inside the ideal mirrors box, then it  
effectively will oscillate  between initial and final state
and so no L state wouldn't be memorized finally, which means that
our IGUS don't function properly.

So we can suppose that this model R  restrictions
 demand that any measurement of MS
don't obstacles $O$ information acquisition and memorization.
In particular any external $E_f$ measurement permitted , but photons reflection
by ideal mirrors to L perturbs $O$ functioning, changing L memorization
conditions and so excluded. 

Note that in this model the role of decoherence differs principally
from Zurek model, where 
 D state collapse is obtained only if 
observer avoid to measure E  operators which are in fact measurable
 $\cite {Zur}$.
This procedure of taking the trace over E states result in
Improper Mixture paradox and was criticized often $\cite {Desp}$.    
  It  demonstrates that like in CH model  at any time moment 
 at least  one IT observable $\hat{B}$ in S-D-E space exists  which
 expectation value $\bar{B}$ coincides with the value for the pure state
and differs from the predicted one for the   mixed state.
So it contradicts with our collapse operational definition.
Moreover it follows that in principle it's possible to restore the
system initial state which contradicts with the irreversibility
 expected for the collapse. In distinction
in our MS model the radiation decoherence is
necessary and inevitable  consequence of  the memorization of  L states.
 IT operator $B$
of the produced field $E_f$ is  principally unmeasurable and
for other operators we have relation (\ref {C22})
, which in operational approach means the collapse.

Note that in our model, even if initial E is vacuum 
our excited system $O$ eventually produces E of new kind -
photon gas. 
In connection with it let's  regard the toy-model
of E production in the measurement. Suppose that some $O$ ( or D)
measuring S emits $N_e>1$ new particles. Then to reconstruct
the complete state we need $N_e$ detectors $D_i$. But each $D_i$
emits also $N_e$ particles, which demands another detectors
ensemble, etc. So in this case the complete MS state depends on
performed measurements and so principally can't be reconstructed.    

\section {Discussion}

In this paper the measurement models which accounts IGUS information
processing and memorization regarded. Real IGUSes are very complicated systems 
with many DFs, but the main quantum effects like superpositions or
decoherence are the same for large and small systems and 
can be studied with the simple models. Obviously any realistic
IGUS can function only in the restricted sector of its atoms
Hilbert space. For example, if to decompose processor chip into
free atoms it wouldn't function. Our CH spin chain model indicates
that IT selfmeasurement can't be performed inside this sector.
In RD model such measurement is incompatible with the photon dissipation
necessary for IGUS functioning and signal memorizing. 
 If we accept as fundamental QM principle :'No observation without
(functioning) observer ' then it follows that some $O$ 
operators are principally unobservable for $O$ $\cite {Sna}$.
So in this approach the collapse problem can be resolved inside standard QM
domain  taking into account observer quantum properties.

Initially the collapse problem was formulated like following:
why in two-slits experiment observer don't see interfering electromagnetic
field radiated from both slits (half and half), but sees the photon
appearing at random from right or left slit ? 
Our suggestive answer is  that observer
perception also obeys QM  laws and due to it the
brain reaction on incoming electromagnetic field described by
entangled state $\Psi_{MS}$ of (1), or (\ref {C1}) if decoherence accounted.
 Perception by brain of the photon coordinate is the local strict
measurement from which in principle can't be reconstructed information
about field interference between left and right slit.

 This CH and RD model results closely connected with operational
definition of collapse  and in its turn relation between strict measurement
and information memorization. We've demonstrated that in all
this cases for $O$ the information discriminating MS pure/mixed
states is principally unavailable. So we can apply     
this collapse definition for $O$ subjective description
of measurement results without contradictions at least for
 models comparison. This operational definition isn't 
sensetive to the 'problem of event' $\cite {Busch}$. 
At this level obtained in chap. 2  observer representation is compatible
with probabilistic description of measurement results, but can't
derive it directly.  

In Relational QM  observers are material local objects which are nonequivalent
in a sense that the  physical world description can be principally different
for them \cite {Rov}. If observer $O$ stops to function (exist), then
some other $O'$ world description principally can't be substituted for
$O$ description.
Due to discussed R restrictions the Univers Hilbert space $H_T$ reduction
to $M_T$ space in $O$ observer representation occurs.
 Note that MS evolution in $H_T$
is formally reversible and
in our models we get irreversible 'subjective' evolution 
observed by $O$ from nonreturnable processes with continuous
spectra like S scattering, excitation of $O$ state and photons radiation.    
Described in chap. 2 $H'_i$ and $M_T$  states manifolds of $O, O'$  
observers can be regarded as  unitarily nonequivalent (UN) representations,
despite their structures needs further mathematical clarification 
$\cite {Ume}$.
Such representations appears also in  nonperturbative Quantum Field 
 Theory   applied already in some microscopic
measurement models $\cite {Fuk}$. In particular
 QED nonperturbative bremsstrahlung  
   model of measurement results in  collapse-like field evolution which 
 reminds our RD model $\cite {May2}$. 

In Everett+brain QM interpretations eq. ($\ref {DD}$)
describes so called observer $O$ splitting identified with state collapse
 $\cite {Whi}$.  In this theory it's  assumed that each
$O$ branch describes the different reality and the state collapse is
phenomenological property of human consciousness. Obviously this
approach has
some common points with our models which deserve further analysis. 
In general all our experimental conclusions are based on human
subjective perception. Assuming the computer-brain perception analogy
in fact means that human signal perception also defined by $\bar{Q}_O$
values. Despite that this analogy looks quite reasonable we can't
give any proof of it.  In our models
in fact  the state collapse have subjective character and 
occurs initially only for single observer $O$, but as was shown
by Rovelli it doesn't results in any contradictions $\cite {Rov}$. 
If it's sensible to discuss any world partition prompted by QM results
it seems to be the border between subject - observer $O$
 which collect information
about surrounding objects S  and objects S which can include other
observer $O'$.

The main conclusion of our paper is that regarded IGUS models
 evidences that QM (and QED) linear evolution can result in
 selfmeasurement restrictions which in operational approach
can be interpreted as the collapse appearance. Altogether we
find independent  effects of three kinds which can induce it : 
$O$ nondestruction, $O$ finite rigid atomic structure,
and decoherence.
 Their mutual relations and influence are unknown, but we don't
expect they will suppress each other. Of them the
decoherence effects and in particular RD seems to us the practically
most important and deserving further detailed study.

\end{document}